\documentclass[journal=nalefd,manuscript=letter,layout=twocolumn]{achemso}

\usepackage{amsmath,graphicx,latexsym,times,color}
\usepackage{setspace}
\usepackage{hyperref}
\usepackage{array}
\usepackage{titlesec}
\usepackage{physics}
\usepackage{color}
\usepackage[T1]{fontenc}
\usepackage[version=3]{mhchem} 

\usepackage{empheq} 

\author{Dongzhe Li}
\email{dongzhe.li@cemes.fr}
\affiliation{CEMES, Universit\'e de Toulouse, CNRS, 29 rue Jeanne Marvig, F-31055 Toulouse, France}

\author{Soumyajyoti Haldar}
\email{haldar@physik.uni-kiel.de}
\affiliation{Institute of Theoretical Physics and Astrophysics, University of Kiel, Leibnizstrasse 15, 24098 Kiel, Germany}

\author{Stefan Heinze}
\affiliation{Institute of Theoretical Physics and Astrophysics, University of Kiel, Leibnizstrasse 15, 24098 Kiel, Germany}
\affiliation{Kiel Nano, Surface, Interface Science (KiNSIS), University of Kiel, 24118 Kiel, Germany}

\title[\texttt{achemso} demonstration]
{Proposal for all-electrical skyrmion detection in van der Waals tunnel junctions}

\newcommand{\V}[1]{\ensuremath{\mathbf{#1}}} 

\let\oldtimes\times  
\renewcommand\times{{\oldtimes}}


\begin{document}
	
	
\begin{tocentry}
\centering
\includegraphics[scale=1]{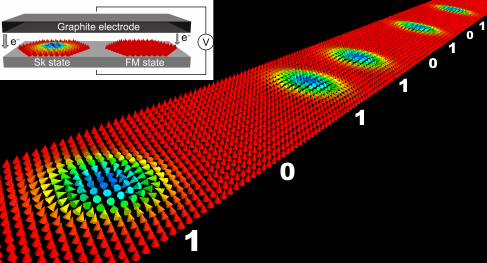}
\end{tocentry}

	\begin{abstract}
A major challenge for magnetic skyrmions in atomically thin van der Waals (vdW) materials is reliable skyrmion detection. Here, based on rigorous first-principles calculations, we show that all-electrical skyrmion detection is feasible in 2D vdW magnets via scanning tunneling microscopy (STM) and in planar tunnel junctions. We use the nonequilibrium Green's function method for quantum transport in planar junctions, including self-energy due to electrodes and working conditions, going beyond the standard Tersoff-Hamann approximation. We obtain a very large tunneling anisotropic magnetoresistance (TAMR) around the Fermi energy for a vdW tunnel junction based on graphite/Fe$_3$GeTe$_2$/germanene/graphite. For atomic-scale skyrmions the noncollinear magnetoresistance (NCMR) reaches giant values. We trace the origin of the NCMR to spin-mixing between spin-up and -down states of $p_z$ and $d_{z^2}$ character at the surface atoms. Both TAMR and NCMR are drastically enhanced in tunnel junctions with respect to STM geometry due to orbital symmetry matching at the interface.~
		\\
	\end{abstract}
	
Magnetic skyrmions \cite{Bogdanov1989} -- topologically stabilized chiral spin structures with size down to the nanometer scale -- have emerged as a promising avenue to realize next-generation spintronic devices~\cite{Nagaosa2013,fert2017magnetic,everschor2018perspective}. Ten years ago, Fert and co-workers first proposed to use skyrmions in a race-track memory in their seminal paper~\cite{fert2013}. Today, many other potential applications of skyrmions are being explored ranging from logic devices to neuromorphic or quantum computing \cite{Luo2018,Back2020,li2021magnetic,Psaroudaki2021}. An essential prerequisite for most applications is reliable electrical detection of individual skyrmions or other topological spin structures.

In ultrathin transition-metal films, skyrmions have been observed directly using spin-polarized scanning tunneling microscopy (STM)~\cite{heinze2011spontaneous,Romming2013}, which is based on the tunneling magnetoresistance (TMR). TMR devices rely on magnetic electrodes, which may perturb the skyrmion state during detection. Skyrmion detection is also possible via the tunneling anisotropic magnetoresistance (TAMR) 
\cite{Bode2002,Gould2004} using non-magnetic STM tips~\cite{heinze2011spontaneous,Bergmann2012,herve2018stabilizing}. However, because TAMR relies purely on spin-orbit coupling (SOC), it is typically too small for device applications. In 2015, 
the noncollinear magnetoresistance (NCMR) has been discovered and proposed for skyrmion detection~\cite{hanneken2015}. NCMR is based on spin-mixing of majority and minority spin channels in a non-collinear spin structure \cite{hanneken2015,crum2015perpendicular,Kubetzka2017,Perini2019}. Since it is not caused by SOC, it can be much larger than TAMR. In STM experiments, the NCMR can be used to observe skyrmions as well as 
domain walls \cite{hanneken2015,Perini2019,muckel2021experimental}. 

An alternative way for all-electrical skyrmion detection is to use the topological Hall effect \cite{Franz2014,Kanazawa2015,maccariello2018electrical,he2023all}. However, such device setups are more difficult to fabricate in terms of device geometries. A simpler solution is to design perpendicular tunnel junctions that can easily integrate skyrmions into conventional semiconductor devices, e.g.,  magnetic tunnel junctions (MTJ) for skyrmions \cite{Penthorn2019,chen2023all}.

More recently, with the discovery of 2D van der Waals (vdW) magnets, a comprehensive study has been performed on the spin transport on 2D magnets in planar junctions \cite{song2018giant,klein2018probing,paudel2019spin2,wang2018tunneling,li2019spin2,min2022tunable}. However, these investigations are restricted to TMR-based devices, i.e., collinear magnetic configurations. There are currently neither experimental nor theoretical works on quantum transport through magnetic skyrmions in 2D vdW magnet planar junctions. Moreover, the Tersoff-Hamann model \cite{Tersoff1985}, which allows explaining the NCMR in STM geometry \cite{hanneken2015,Perini2019}, is questionable if one aims to explore planar tunnel junction devices, where the detecting electrodes are not sharp tips and there is no vacuum gap as in STM experiments.

Here, we demonstrate based on first-principles calculations
that the NCMR and the TAMR 
can be very large in 2D vdW magnets,
suggesting the possibility of
all-electrical detection of magnetic skyrmions.
We study both the regime of tunneling across a vacuum gap as applicable to STM as well as tunnel junction devices with non-magnetic electrodes. 
We predict a giant NCMR for 
atomic-scale skyrmions in a
graphite/Fe$_3$GeTe$_2$/germanene/graphite tunnel junction.
The NCMR results from spin mixing of spin-up and -down 
states of $p_z$ and $d_{z^2}$ character at the surface atoms.
Due to orbital symmetry matching at the interface 
it is enhanced 
by orders of magnitude in the tunnel device with respect to that obtained 
in STM geometry. We find a similar enhancement of the TAMR.
These magnetoresistance values are at least one order of magnitude larger than those reported so far for conventional transition-metal interfaces \cite{hanneken2015,Perini2019}.
Our work also shows the importance
of employing non-equilibrium Green's functions (NEGF) for quantum transport 
of non-collinear spin states in tunnel junctions, where the 
TH approximation proves inadequate. 

\begin{figure*}[t]
	\centering
	\includegraphics[width=0.66\linewidth]{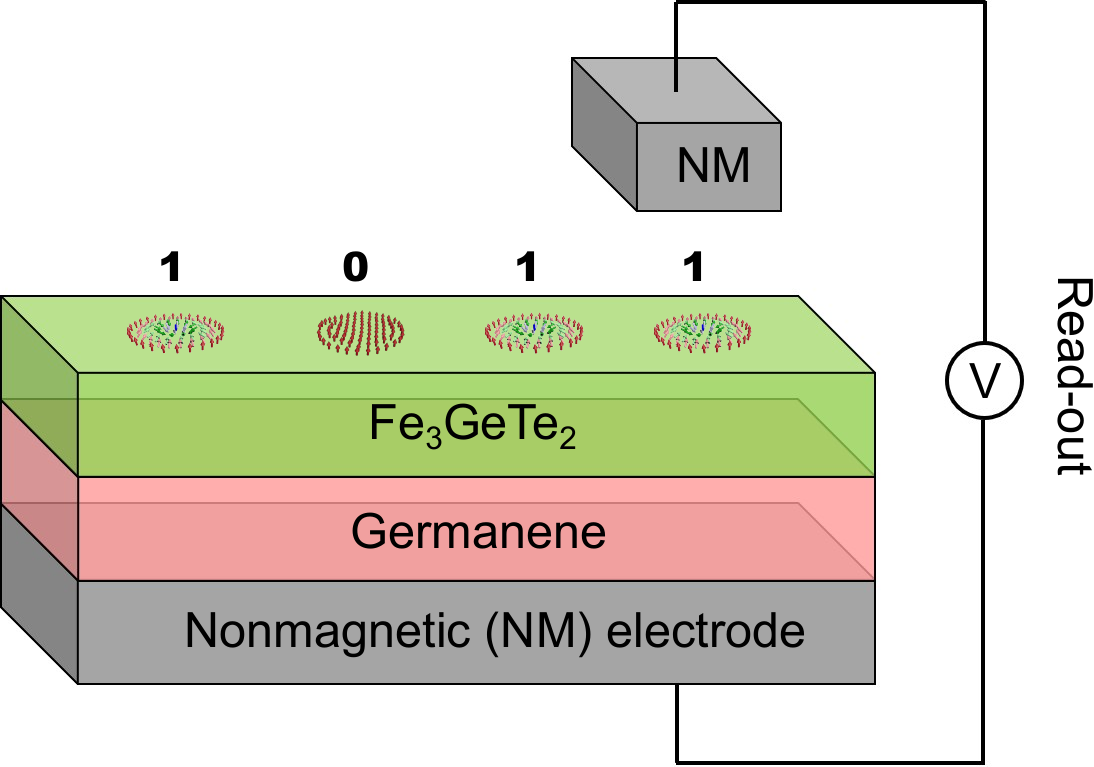}\\
	\caption{Schematic representation of the proposed vertical tunnel junctions with nonmagnetic electrodes (TJ-NM) for electrical read-out of skyrmions in 2D vdW magnets, e.g., in racetrack memory. The skyrmions are stabilized at the Fe$_3$GeTe$_2$/germanene vdW heterostructure. Reading data from the skyrmion pattern is accomplished all electrically based on NCMR.}
	\label{FGT_ge_structure} 
\end{figure*}

Fig.~\ref{FGT_ge_structure} shows a sketch of our proposal for all-electrical skyrmion detection, consisting of Fe$_3$GeTe$_2$/germanene (FGT/Ge), a representative 2D vdW heterostructure, sandwiched between two nonmagnetic electrodes, namely tunnel junctions with nonmagnetic electrodes (TJ-NM). A key difference between a TJ-NM and the widely used MTJ setup is that we do not rely on an external magnetic field for device operations. We demonstrate that this setup allows detection of the difference in perpendicular current flow for isolated skyrmions and their ferromagnetic surrounding, enabling an all-electrical detection scheme.

\begin{figure*}[t]
	\centering
	\includegraphics[width=1.0\linewidth]{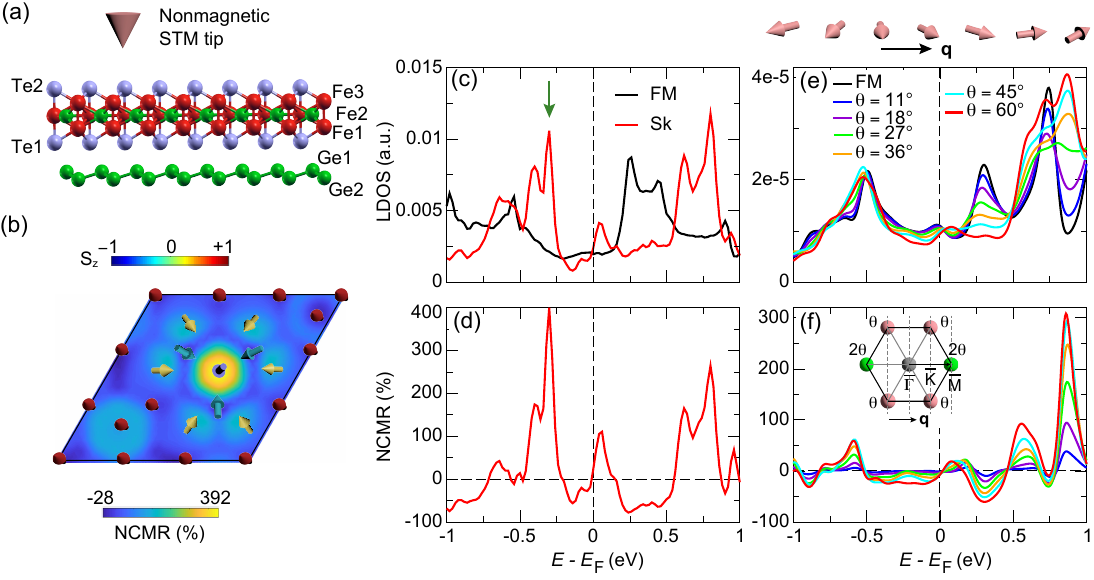}
	\caption{Calculated NCMR in STM geometry using the TH approximation. (a) Schematic plot of an STM experiment on FGT/Ge with a non-magnetic tip. (b) Spin structure of the nanoscale Néel-type skyrmion stabilized in strained FGT/Ge~\cite{Dongzhe2022_fgt} and the NCMR map calculated at a distance of 3 {\AA} from the surface and an energy of $E=E_{\text{F}}-0.3$~eV. (c) Vacuum LDOS for the FM state (black) and the Sk state (red). The green arrow marks the energy at which the NCMR map in panel (b) has been plotted. (d) NCMR calculated from the LDOS of the FM and Sk states shown in panel (c). (e) LDOS in the vacuum calculated for spin spiral states at 3 \AA~above FGT/Ge for various nearest-neighbor angles $\theta$. (f) NCMR calculated from the LDOS of the FM and spin spiral states shown in panel (e). The inset shows the hexagonal atomic lattice with the spin spiral vector $\V{q}$ along the $\overline{\Gamma} - \overline{\text{K}} - \overline{\text{M}}$ direction of the 2D Brillouin zone and the angle $\theta$ between spins on neighboring sites. 
	}
	\label{ncmr_ss} 
\end{figure*}

We consider FGT/Ge (Fig.~\ref{ncmr_ss}a) as a representative 2D vdW magnet heterostructure and skyrmion platform for several reasons. First, there is strong experimental evidence for magnetic skyrmions in vdW FGT heterostructures \cite{ding2019observation,wu2020neel,Park2021,wu2021van}. Second, FGT has a very high Curie temperature, which can vary from 150 K to 220 K depending on Fe occupancy \cite{deng2018gate,li2018patterning}. Third, recent theoretical work based on first-principles calculations and atomistic spin simulations predicts that magnetic interactions in FGT/Ge are highly tunable by strain \cite{Dongzhe_prb2023}, leading to stabilize nanoscale skyrmions with diameters down to a few nanometers \cite{Dongzhe2022_fgt}. Finally, the properties of
atomically thin vdW layers can be controlled by external stimuli, which makes them ideal skyrmion platforms from a materials design perspective.

Electronic structure and quantum transport calculations were carried out using \textsc{QuantumATK} \cite{Smidstrup_2020}, which uses the non-equilibrium Green function (NEGF) \cite{Mads2002} formalism combined with noncollinear density functional theory (DFT) \cite{Gross2013PRL}. Additionally, we used \textsc{Fleur} \cite{fleur_v26} to perform spin spiral calculations based on the generalized Bloch theorem \cite{Kurz2004}. Computational details are given in Section I in Supporting Information.

\noindent{\textbf{NCMR in STM geometry.}} 
We start our discussion with the NCMR effect in STM geometry (Fig.~\ref{ncmr_ss}a) using the TH approach~\cite{Tersoff1985}. Due to the large computational cost, we used the smallest Néel-type skyrmion, which fits into a 
$(3 \times 3)$ FGT/Ge supercell (Fig.~\ref{ncmr_ss}b). 
The angle between adjacent spins is $60^\circ$ and the
estimated skyrmion diameter is around 
1.2~nm. This is an atomic-scale skyrmion for which quantum
transport calculations based on the NEGF formalism are feasible.
 Note, that only slightly larger skyrmions were found to be stable in atomistic spin simulations using parameters from DFT \cite{Dongzhe2022_fgt}. 
According to the TH model \cite{Tersoff1985}, the differential conductance, $dI/dU$, in an STM experiment is given by
\begin{equation}\label{TH}
	dI/dU (\mathbf{R}_{\rm T},U) \propto n (\mathbf{R}_{\rm T},E_{\rm F}+eU)
\end{equation}
where $\mathbf{R}_{\rm T}$ is the tip position, $U$ is the bias voltage, and $n(\mathbf{r},E)$ is the local density of states (LDOS) of the sample evaluated in the vacuum a few {\AA} above the surface.
Even if the STM tip is non-magnetic, the LDOS and thereby the obtained $dI/dU$ signal can be sensitive to the local spin texture due to NCMR \cite{hanneken2015}, defined for a skyrmion (Sk) in the ferromagnetic (FM) background as $\text{NCMR} = \frac{dI/dU_{\rm Sk}-dI/dU_{\rm FM}}{dI/dU_{\rm FM}}$. 

The NCMR signal varies locally above the Néel-type skyrmion (Fig.~\ref{ncmr_ss}b) and reaches a maximum value of about 392~\% at the skyrmion core. In a ring around the core the value drops to about $-$21~\% and rises again to about 90~\% as one moves further from the center. As expected, the NCMR contrast becomes smaller as one approaches the edge of the skyrmion since the effective non-collinearity is reduced close to the FM environment. 

From the energy-resolved vacuum LDOS of the skyrmion core vs.~the FM environment (Fig.~\ref{ncmr_ss}c) evaluated at the skyrmion core above the Te atom, we observe the large NCMR effect of varying positive and negative sign at various energies around $E_{\text{F}}$ (Fig.~\ref{ncmr_ss}d). A similar energy dependence of the NCMR is found at other STM tip positions, and we find that the SOC contribution to the NCMR is rather small. We have also performed calculations for a skyrmion in a $(4 \times 4)$ FGT/Ge supercell with angles of $45^\circ$ between adjacent spins and
a skyrmion diameter of around 1.6~nm, which exhibits an NCMR effect of a similar order of magnitude and lateral variation (see Fig.~S2 and Fig.~S3 in Supporting Information). Additionally, we have checked the effect of the size of the FM background on NCMR for the same skyrmion size (see Fig.~S4 in Supporting Information); the magnitude and energy dependence of the NCMR do not vary significantly if we increase the FM background area.

To go beyond the nanoscale skyrmion (Fig.~\ref{ncmr_ss}b) and to vary the period of the noncollinear spin structure on a larger scale, we locally approximate the electronic structure in a skyrmion by that of a homogeneous spin spiral state \cite{Kurz2004}. This approach can explain the experimentally observed NCMR effect of skyrmions and domain walls in ultrathin films \cite{hanneken2015,Perini2019}. Note, that spin-polarized STM was used to resolve domain walls in FGT \cite{Yang_2022,Trainer2022}. 

Fig.~\ref{ncmr_ss}e shows the vacuum LDOS calculated about 3~{\AA} above FGT/Ge for spin spirals of different periods. As the spin spiral rotating angle $\theta$ varies, one observes significant changes in the height and position of the peaks in the vacuum LDOS (Fig.~\ref{ncmr_ss}e). 
We find a prominent peak at about 0.25~eV above $E_{\rm F}$, which quickly decreases with rising non-collinearity, and one at about 0.75~eV above $E_{\rm F}$, which shows a more complex change. In contrast, the peak at about $0.5$~eV below $E_{\rm F}$ displays only a small shift and broadening.

The corresponding NCMR (Fig.~\ref{ncmr_ss}f) calculated for various spin spiral states 
\footnote{The NCMR spectrum for spin spirals (SS) is calculated by $\text{NCMR} = \frac{dI/dU_{\rm SS}-dI/dU_{\rm FM}}{dI/dU_{\rm FM}}$}
shows a large negative value around $E_{\rm F}+0.25$~eV up to about 80~\% due to the vanishing peak (cf.~Fig.~\ref{ncmr_ss}e). At larger energies, we find two peaks in the NCMR of up to
about 100\% and 300\%. Note that the 
spin spiral periods, 
given by $\lambda = 2\pi/\V{|q|}$, are 
about 6.5, 3.9, 2.6, 1.9, 1.6 and 1.2~nm for 
angles of 
$\theta$ = 11$^{\circ}$, 18$^{\circ}$, 27$^{\circ}$, 36$^{\circ}$, 45$^{\circ}$ and 60$^{\circ}$, respectively. The NCMR spectrum of the spin spiral state with the shortest period, i.e.~largest angle $\theta=60^\circ$ (red curve in Fig.~\ref{ncmr_ss}f), is similar in sign and order of magnitude to that obtained for the skyrmion (Fig.~\ref{ncmr_ss}d) since the rotating angle between neighboring spins in our atomic-scale skyrmion (Fig.~\ref{ncmr_ss}b) is close to $60^{\circ}$. An exception is the large NCMR peak at about 0.3~eV below $E_{\rm F}$ for skyrmions, which is missing for spin spirals.

\begin{figure*}[t]
	\centering
	\includegraphics[width=1.0\textwidth]{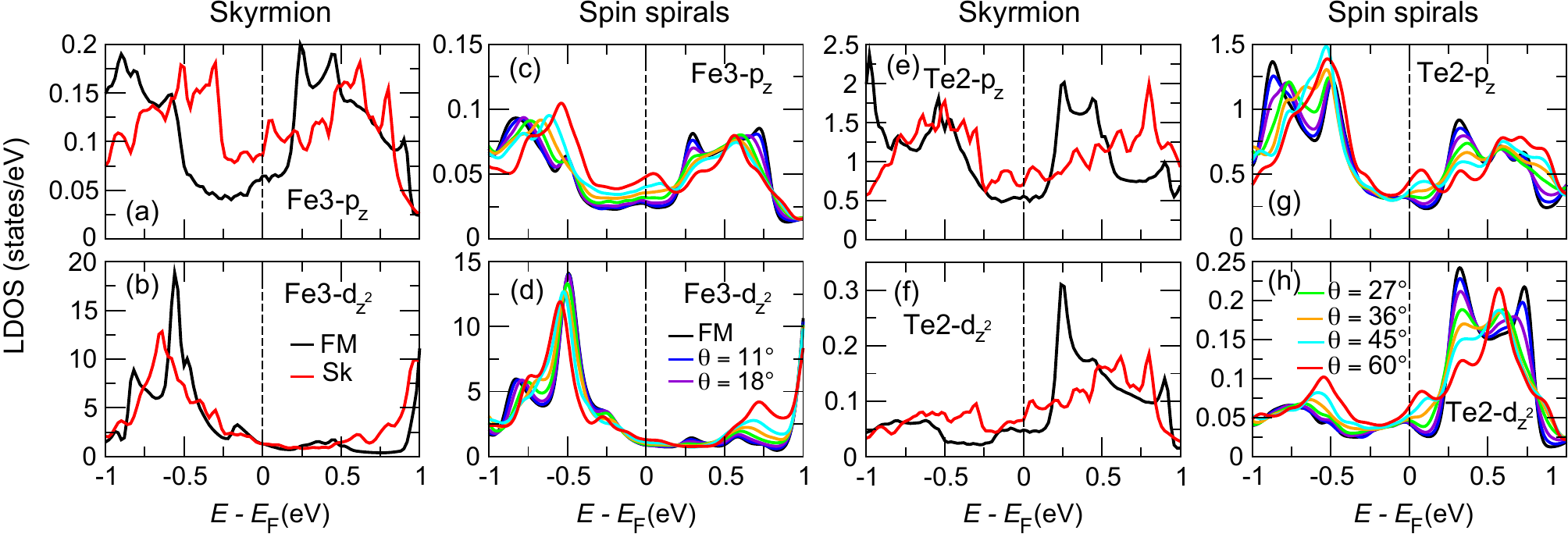}\\
	\caption{LDOS of (a) $p_z$ and (b) $d_{z^2}$ character in the FM (black) and in the skyrmion state (red) for the Fe3 (surface) atom (cf.~Fig.~2a) in the FGT/Ge heterostructure. Note, that the sum of spin-up and -down states is shown. (c-d) as (a-b) for the spin spiral state for various angles $\theta$ between adjacent magnetic moments. (e-h) as (a-d) for the Te2 (surface) atom.  
	}
	\label{pdos_compare} 
\end{figure*}

The origin of the NCMR for skyrmions can be understood by analyzing the electronic structure. We find the vacuum LDOS (Fig.~\ref{pdos_compare}) to be dominated by the $p_z$ and $d_{z^2}$ states of Te2 and Fe3 since these orbitals exhibit the slowest decay into the vacuum. The characteristic changes of the vacuum LDOS for the skyrmion vs.~FM state (Fig.~\ref{ncmr_ss}c), i.e.~the shifted double peak structure above $E_{\rm F}$ and the extra peak at about $E_{\rm F}-0.3$~eV for the skyrmion, are clearly visible in the Fe3-$p_z$ LDOS (Fig.~\ref{pdos_compare}a) and in the Te2-$p_z$ and Te2-$d_{z^2}$ LDOS (Fig.~\ref{pdos_compare}e,f).

For the spin spiral states, we obtain similar conclusions. The variation of the angle $\theta$ between adjacent spins leads to a gradual change of the peaks, e.g., the decreasing peak height in the vacuum LDOS at about $E_{\rm F}+0.25$~eV (Fig.~\ref{ncmr_ss}e).
This effect also shows in the Fe3-$p_z$ LDOS (Fig.~\ref{pdos_compare}c) and in the Te2-$p_z$ and Te2-$d_{z^2}$ LDOS (Fig.~\ref{pdos_compare}g,h). In the LDOS of Fe3-$d_{z^2}$ (Fig.~\ref{pdos_compare}d) we notice a shift and decrease of the peak at $E_{\rm F}-0.5$~eV which explains the change of the vacuum LDOS at this energy (Fig.~\ref{ncmr_ss}e). For more detailed $p_z$- and $d_{z^2}$-orbital resolved NCMR from the Fe3 and Te2 atoms, see Fig.~S5 and Fig.~S6 in Supporting Information. Note, that the variations of the orbital decomposed LDOS at the Fe3 and Te2 atoms is also similar for the skyrmion and spin spiral state with the largest angle, i.e.~$\theta=60^\circ$ (red). 

\begin{figure}[t]
	\centering
	\includegraphics[width=1.0\linewidth]{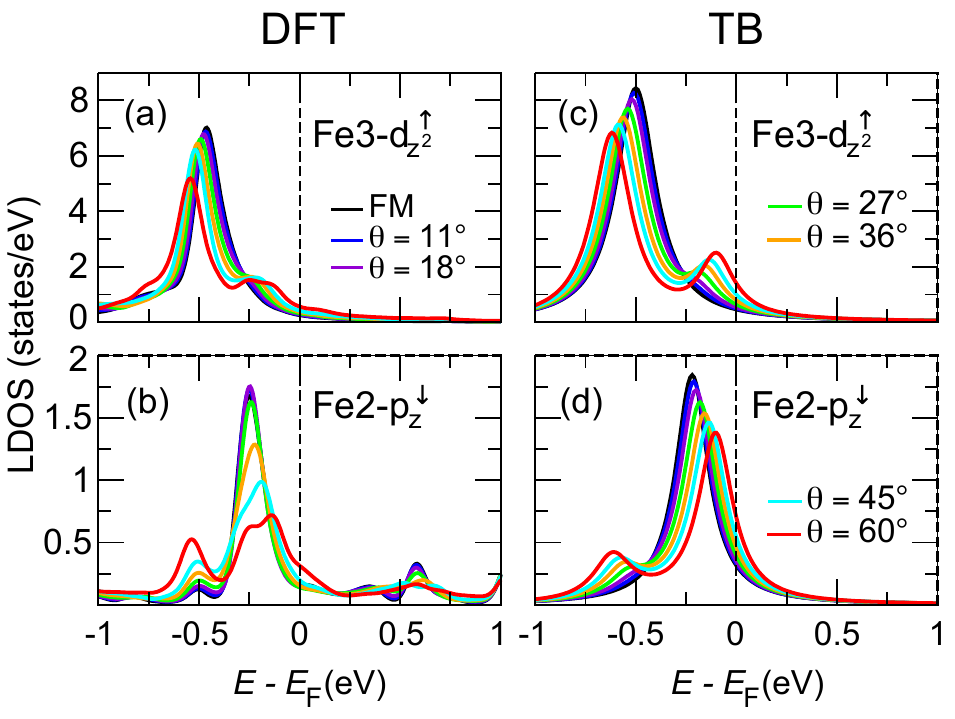}
	\caption{\label{fig:TB} 
     NCMR mechanism by spin-mixing in FGT/Ge. LDOS of Fe3-$d_{z^2}^{\uparrow}$ (a) and Fe2-$p_{z}^{\downarrow}$ (b) states as obtained from DFT calculations for spin spiral states with an angle $\theta$ between adjacent spins. (c-d) The same as in (a-b) but using a two-level TB model proposed in Ref.~\cite{Perini2019}. (See Section I in Supporting Information for more details.)
 }
       \label{TB-model} 
\end{figure}

The physical mechanism of NCMR can be elucidated through the spin mixing effect, mainly driven by interlayer hopping between Fe3-$d_{z^2}^\uparrow$ and Fe2-$p_{z}^\downarrow$, as well as intralayer hopping between Te2-$p_{z}^\uparrow$ and Te2-$p_{z}^\downarrow$. The shift of the $d_{z^2}^\uparrow$ peak at about 0.5~eV below $E_{\rm F}$ at the Fe3 atom can be attributed to spin mixing with $p_z^\downarrow$ state peak at $E_{\rm F}-0.25$~eV of the Fe2 atom (Fig.~\ref{TB-model}a-b). The spin mixing leads to a shift and splitting of the peaks in both LDOS, and this effect is reproduced and understood by a two-level tight binding (TB) model proposed in Ref.~\cite{Perini2019}, as shown in Fig.~\ref{TB-model}c-d. For the $p_z$ states at the Te2 atom a similar effect is found for the states above $E_\text{F}$ (see Fig.~S7 in Supporting Information). Note, that these changes are directly reflected in the vacuum LDOS (Fig.~\ref{ncmr_ss}e) and can explain the NCMR spectrum above $E_{\rm F}$ (Fig.~\ref{ncmr_ss}f). See Figs.~S8-S10 in Supporting Information for more detailed LDOS of the Te2, Fe3, and Fe2 atoms.

\noindent{\textbf{NCMR in tunnel junctions.}} 
To properly address NCMR in tunnel junctions, including self-energy due to electrodes and working conditions, one has to calculate the nonequilibrium charge/spin density using the NEGF formalism \cite{Mads2002}, going beyond the TH approximation. The transmission function is calculated by NEGF as

\begin{equation}
	\label{eq:negf}
	T(E)=\Tr[\V{\Gamma_{\text{L}}}\V{G}\V{\Gamma_{\text{R}}}\V{G}^{\dagger}]
\end{equation}
where $\V{G}$ is the retarded Green's function of the central region, and $\V{\Gamma_{\text{L}}}$/$\V{\Gamma_{\text{R}}}$ are matrices describing its coupling to left/right semi-infinite electrodes. 

\begin{figure}[t]
	\centering
	\includegraphics[width=1.0\linewidth]{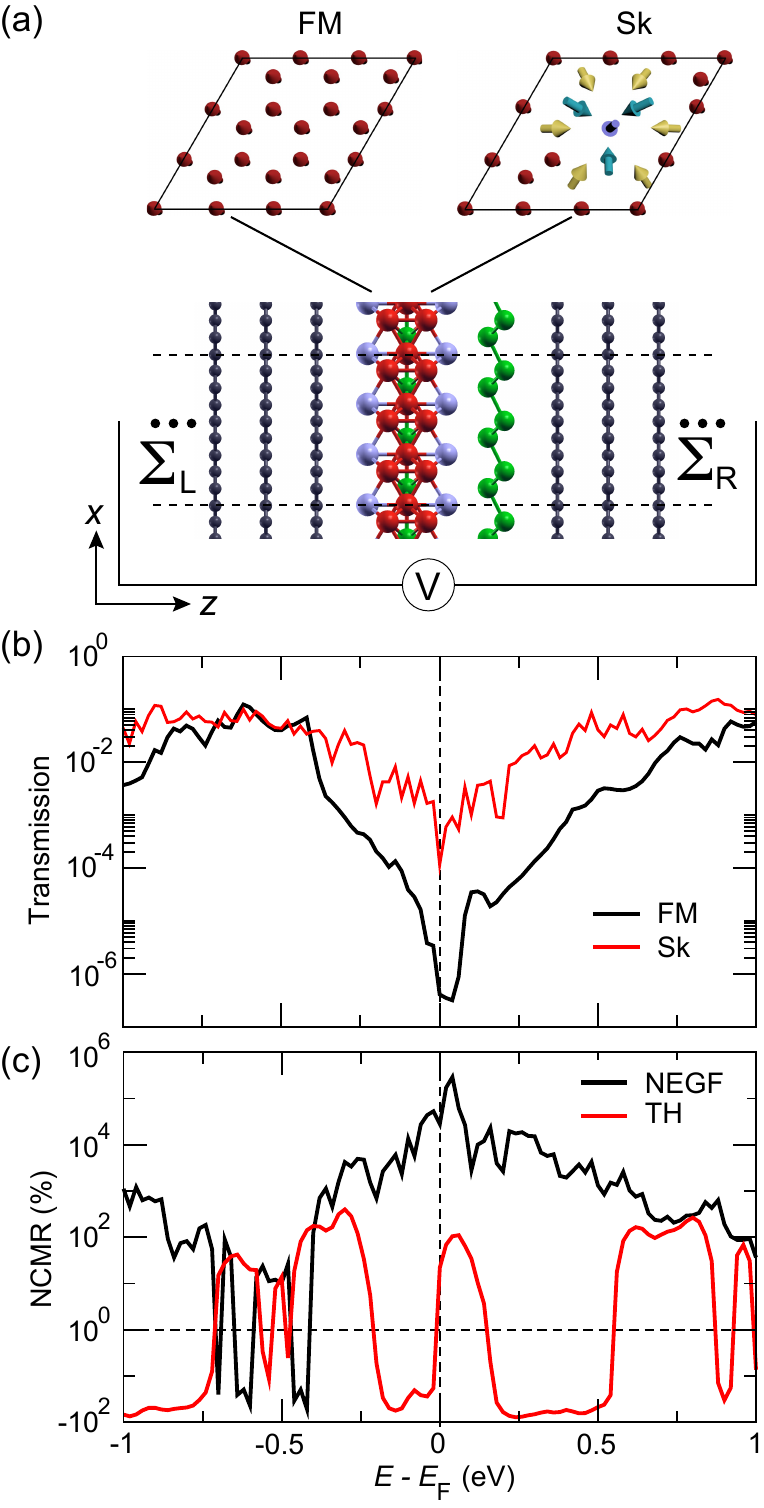}
	\caption{Calculated NCMR in tunnel junctions using the NEGF formalism. 
 (a) Side view of the atomic structure of the graphite/FGT/Ge/graphite junction device used for measuring current difference in the FM and Sk state. (b) Transmission functions for the FM state (black), and the (3 $\times$ 3) Néel-type Sk state (red). (c) The corresponding NCMR (black), defined as $\text{NCMR}=[T_{\text{Sk}}(E)-T_{\text{FM}}(E)]/T_{\text{FM}}(E)$, calculated by NEGF. The NCMR obtained in the TH approximation (red) is shown for comparison (same data as in Fig. \ref{ncmr_ss}d).
	}
	\label{NCMR_NEGF}
\end{figure}

We propose to consider a TJ-NM created by the graphite/FGT/Ge/graphite junction to detect a skyrmion by all-electrical means (Fig.~\ref{NCMR_NEGF}a-b). Note, that we used a FGT/Ge $(1 \times 1)$ lattice strained by $-3$~\% as a fixed layer, and a $\sqrt{3}$ $\times$ $\sqrt{3}$ in-plane unit cell of graphite is matched to it. Fig.~\ref{NCMR_NEGF}c shows zero-bias transmission functions through an atomic-scale skyrmion (spin structure as in Fig.~\ref{ncmr_ss}b) and the FM state for the tunnel junction. The junction exhibits an insulating feature for the FM state with a clear dip at $E_{\text{F}}$ due to the orbital symmetry matching effect between C-$p_z$ and Te-$p_z$ orbitals at the interface. Note that the hybridization effect between FGT/Ge and the graphite electrodes is small due to the weak vdW interaction (see Fig.~S11 in Supporting Information).

The transmission for the skyrmion state (Fig.~\ref{NCMR_NEGF}b) exhibits much larger values in a broad energy range, leading to an extremely large NCMR of more than $10,000$~\% near $E_{\text{F}}$ (Fig.~\ref{NCMR_NEGF}c). This value is at least two orders of magnitude higher than that observed for transition-metal interfaces \cite{hanneken2015,crum2015perpendicular,Perini2019,Richarz2019}. 
The NCMR calculated by NEGF differs significantly from that obtained in the TH model (Fig.~\ref{NCMR_NEGF}c),
especially around $E_\text{F}$. The NCMR obtained via the TH approach changes sign several times in the considered energy range and reaches a maximum of about 400\% at 0.3 eV below $E_{\text{F}}$. In contrast, the NCMR calculated by NEGF is positive in almost the entire energy range and between $1,000$ and $10,000$~\% in a wide range around $E_{\text{F}}$, reaching a maximum value of about $100,000$~\%. This demonstrates that the proposed TJ-NM is an ideal platform for all-electrical detection of skyrmions. 

The extremely large NCMR observed in TJ-NM stems from the interplay of the symmetry of electronic states and the spin mixing effect. Due to orbital symmetry matching at the interface of FGT/Ge and the graphite electrodes, C-$p_z$ mainly couples to $p_z$ orbitals of the surface atoms of FGT/Ge (Te2, Fe3, and Ge2) (see Fig.~S12 and S13 in Supporting Information). The shape of the LDOS at the C atoms in combination with the low $p_z$ LDOS at the Te2, Fe3, and Ge2 atoms near $E_{\rm F}$ can explain the v-shaped transmission functions (see Fig.~S14 in Supporting Information). Due to spin mixing in the skyrmion state the $p_z$ LDOS rises around $E_{\rm F}$ at the Te2 and Fe3 atoms as discussed above -- an effect that is also clearly visible in the device setup. This explains the enhanced transmission for the skyrmion state in the vicinity of $E_{\rm F}$. At energies above $E_{\rm F}+0.2$~eV, the NCMR remains positive (Fig.~\ref{NCMR_NEGF}c), while the $p_z$ LDOS of the Te2 and Fe3 surface atoms is higher in the FM state between 0.2 and 0.5~eV above $E_{\rm F}$. Accordingly, the NCMR within the TH model is negative since the tunneling current only depends on the vacuum LDOS that is dominated by surface localized states and not on the coupling of the states to the bulk states. In contrast, for a large transmission obtained by NEGF, the contributing states need to extend through the entire FGT/Ge layer. The higher transmission in the skyrmion and positive NCMR above $E_{\rm F}+0.2$~eV indicate that the $p_z$ states, which leads to a large LDOS between $E_{\rm F}+0.2$~eV and $E_{\rm F}+0.5$~eV in the FM state, contribute little to the current due to their strong localization at the surface atoms. This interpretation is supported by the spatial localization visible in the LDOS map of FGT/Ge (see Fig.~S15 in Supporting Information). Note, that a full understanding of the transmission function based only on the LDOS is not feasible. Nevertheless, our analysis illustrates that for a device setup, both interfaces to the electrodes need to be considered and can lead to large enhancements of the NCMR.

We have also calculated the TAMR of our tunnel junction (Fig.~S16 in Supporting Information).
The obtained value of about 200\% near $E_{\text{F}}$ is much smaller than the
NCMR since the TAMR originates from SOC. However, the value is still more than one order of magnitude larger than the TAMR reported in ultrathin films \cite{Bode2002,Bergmann2012,crum2015perpendicular,Kubetzka2017,Richarz2019} and similar to that reported for molecular junctions \cite{li2015giant,Rakhmilevitch2016,Li_PRR2020,Otte2015}, showing again the high promise of the proposed TJ-NM for all-electrical skyrmion detection.
 
Note that due to an extremely high computational cost, we are restricted to an atomic-scale skyrmion  
in our quantum transport calculations for the device setup. 
The obtained NCMR (Fig.~\ref{NCMR_NEGF}c)
applies to a state of densely packed skyrmions or a tunnel junction with a cross-section of the size of the $(3 \times 3)$ 2D unit cell. An increase in the cross-section will lead to a reduction of the NCMR values given here. 
Skyrmions with a larger diameter, which can be realized experimentally,
will also reduce the NCMR as shown by our spin spiral calculations for the STM setup (cf.~Fig.~\ref{ncmr_ss}(e,f)).

However, the key message of our quantum transport calculations is that the NCMR signal is greatly enhanced in a tunnel device with respect to STM geometry (Fig.~\ref{NCMR_NEGF}c)
due to orbital symmetry matching -- an electronic effect due to hybridization
at the interface which is expected as well
for nanoscale skyrmions. This expectation is supported by the calculation 
of the TAMR (Fig.~S16 in Supporting Information), which is also 
enhanced by up to one order of magnitude in planar tunnel junctions compared 
to that in STM geometry due to orbital symmetry matching.

To conclude, we suggest planar devices built from 2D vdW heterostructures such as the graphite/FGT/Ge/graphite tunnel junction 
considered as an example in our work
as ideal platforms for reliable all-electrical skyrmion detection down to the atomically thin limit. Very large NCMR and TAMR are observed 
in our calculations
near the Fermi energy, more than one order of magnitude higher than those
reported for transition-metal interfaces. The physical mechanism of the NCMR is explained by the interplay between the spin mixing and orbital symmetry matching effects at the interface. Our work highlights the crucial importance of using the NEGF approach for quantum transport on noncollinear spin structures in tunnel junctions, going beyond the TH model. Our proposal opens a new route to realize skyrmion racetrack memories based on atomically thin vdW materials with full-electrical writing and reading.
	
\textbf{Acknowledgement:} This study has been supported through the ANR Grant No. ANR-22-CE24-0019. This study has been (partially) supported through the grant NanoX no.~ANR-17-EURE-0009 in the framework of the ``Programme des Investissements d'Avenir''. Financial support from the Deutsche Forschungsgemeinschaft (DFG, German Research Foundation) through SPP2137 ``Skyrmionics'' (project no.~462602351) is gratefully acknowledged. We acknowledge CALMIP (Grant 2023-[P21008]) and the North-German Supercomputing Alliance (HLRN) for providing HPC resources. We thank M. A. Goerzen and H. Schrautzer for helpful discussions.
		
\textbf{Author contributions:} D.L. and S. Heinze conceived the project. D.L. and S. Haldar performed spin spiral calculations in 2D van der Waals heterostructures. D.L. performed first-principles quantum transport simulations through individual skyrmions in tunnel junctions. All authors participated in discussions, analyzed the results, and agreed on the contents included. D.L.~and S. Haldar prepared the figures. D.L.~and S Heinze~wrote the manuscript with the input of all authors.

	\begin{suppinfo}
		The Supporting Information is available free of charge at \url{https://pubs.acs.org/}. Computational details, STM tip position and SOC effects on NCMR, the influence of skyrmion size on NCMR in FGT/Ge, the effect of the FM area size on NCMR, orbital-resolved NCMR by the Fe3 and Te2 surface atoms, the spin-mixing effect induced by the Te2 atom, spin- and orbital-resolved LDOS for the Fe3, Fe2 and Te2 atoms, the orbital hybridization effect at the interface between FGT/Ge and graphite electrodes, atom and orbital-resolved LDOS of the FGT/Ge interface for the FM and Sk states, LDOS map, as well as the TAMR calculated by TH in STM geometry and by NEGF in the tunnel junction.
		
	\end{suppinfo}
	
	\bibliography{Bibliomanuscript}

\end{document}